\documentclass[preprint,aps,prb]{revtex4}
\usepackage{latexsym}
\input{epsf}
\def\sc{\scriptsize}

\begin{document}
\draft
\title{Computational leakage: Grover's algorithm with imperfections}

\author{Pil Hun Song and Ilki Kim\hspace{.5mm}\cite{AUTH}}
\affiliation{\vspace*{1.2ex}Department of Electrical and
         Computer Engineering, North Carolina State University, Raleigh,
         NC 27695-8617, U.S.A.\vspace*{1.0ex}}
\date{\today}
%

\begin{abstract}
We study the effects of dissipation or leakage on the time
evolution of Grover's algorithm for a quantum computer. We
introduce an effective two-level model with dissipation and
randomness (imperfections), which is based upon the idea that
ideal Grover's algorithm operates in a 2-dimensional Hilbert
space. The simulation results of this model and Grover's algorithm
with imperfections are compared, and it is found that they are in
good agreement for appropriately tuned parameters. It turns out
that the main features of Grover's algorithm with imperfections
can be understood in terms of two basic mechanisms, namely, a
diffusion of probability density into the full Hilbert space and a
stochastic rotation within the original 2-dimensional Hilbert
space.
\end{abstract}


\maketitle
%

Recently, quantum computing has emerged as one of the most
challenging fields of physics both for theoreticians and
experimentalists (see Ref.~\onlinecite{steane} for a review). At
the core of the theoretical side, a few quantum algorithms are now
available, which can solve a certain class of problems faster than
any available classical counterparts: for example, S{\sc{HOR}}'s
algorithm \cite{shor} factorizes a given large number $N$ at $\sim
(\log N)^2$ time steps with an exponential speed-up. Using
G{\sc{ROVER}}'s algorithm (GA),\cite{grover} one can find a
specific item on a long list of size $N$ at $\sim \sqrt{N}$ time
steps, which is a considerable gain in speed as compared with
$\sim N$ in classical algorithms.

These quantum algorithms operate perfectly only on ideal quantum
computers. On the other hand, a certain amount of dissipation or
uncontrolled coupling to the environment is clearly inevitable on
real quantum computers. For example, any deviation from ideal
operation in quantum gates, which may result from various origins,
including fluctuation in the excitation energies of two-level
systems (qubits), can be considered as ``imperfections''. The
imperfections will affect the efficiency of a quantum computer,
and the operability of a given quantum algorithm may break down to
the point of losing its advantage over a classical counterpart.
Therefore, it is of vital importance to have a sound picture of
how an error due to the presence of imperfections evolves in
quantum algorithms. Obviously, a reasonable picture of the basic
mechanisms given by the imperfections will be very crucial in
constructing an appropriate quantum error correction
method.\cite{cal,steane2,preskil,gottes}

In general, the quantum state in a quantum computer is essentially
a many-body (or network) state, the time evolution of which is
delicately controlled by a given quantum algorithm. From such a
point of view, the study of imperfection effects on quantum
algorithms would belong to a more general research field which
investigates disorder effects on the dynamics of a many-body
state. Their exact treatment is actually a complicated subject,
and only a few results have been obtained giving either general
frameworks for understanding the effects or general methodologies
for calculation.

There exist several theoretical, mainly numerical, investigations
in this direction. The main stress has been given, from a
practical point of view, on the stability of quantum algorithms
with respect to the presence of imperfections. C{\sc{IRAC}} and
Z{\sc{OLLER}}\cite{zoller} reported that the operability of
quantum computing is rather safe against disorders available in
the quantum F{\sc{OURIER}} transform process. In
Refs.~\onlinecite{paz} and \onlinecite{zurek}, the disorder effect
in S{\sc{HOR}}'s algorithm applied to the factorization of the
number 15\, was studied and by using the fidelity being defined as
the square of the overlap of the actual quantum state with the
ideal one, it was found that the operability of the S{\sc{HOR}}'s
algorithm can be destroyed due to a very small strength of the
disorder in the modular exponentiation part.\cite{zurek} More
systematic results have recently been obtained in
Ref.~\onlinecite{song} from the study of quantum computing of
quantum chaos and imperfection effects: by considering the
presence of imperfections in the quantum F{\sc{OURIER}} transform,
it was obtained that the imperfection strength scales polynomially
with the number of qubits for the inverse participation ratio
(IPR), which measures the strength of localization of quantum
state and plays a role of the fidelity in Ref.~\onlinecite{zurek}.
Nevertheless, it still remains at a primitive stage regarding an
understanding of basic mechanisms carried by the imperfections in
quantum algorithms. So far, the main policy has been simply to
watch a deviation of the quantum state from the ideal one and to
analyze its parameter dependence.

In this paper, we investigate the time evolution of a state
governed by G{\sc{ROVER}}'s algorithm with imperfections, with a
main emphasis on an understanding of interplay of the
imperfections with the algorithm operator. Based on the idea that
the ideal GA operates in an effective 2-dimensional H{\sc{ILBERT}}
space, a stochastic two-level model with dissipation will be
introduced, and then its simulation results will be compared to
those of the GA with imperfections, which operates in a larger
relevant H{\sc{ILBERT}} space resulting from the presence of the
imperfections. They are in a good agreement via an appropriate fit
of parameters. An analytic solution of the two-level model is
given with some modification and provides a comprehensive picture
of imperfection effects on the GA.

Let us begin with a brief sketch of the GA. The final goal is to
identify $|j\rangle$ (target state) among $N = 2^{n_q}$ quantum
states, where $n_q$ is the number of qubits. Initially, the state
of quantum register is prepared as a superposition of all states
with the same amplitude. The GA may be broken up into two steps:
(i) rotation of phase of $|j\rangle$ by $\pi$ and (ii) application
of a diffusion operator $D$ which is defined, in matrix form, as
$D_{kl}=-\delta_{kl}+2/N$ with $k,l = 0,1, \cdots, N-1$, and
$\delta_{kl}$ denoting the K{\sc{RONECKER}} delta. The step (ii)
is achieved by applying the H{\sc{ADAMARD}} operation to each
single qubit and then performing a conditional phase shift on the
computer with every computational basis state except $|k=0\rangle$
receiving a phase shift of $-1$ followed by the second
H{\sc{ADAMARD}} operation to each single qubit. Then, the quantum
state during time evolution can be expressed as \cite{boyer}
\begin{equation}
|\Psi(\vartheta)\rangle\; =\; \sin \vartheta\,|j\rangle\, +\,
\frac{\cos \vartheta}{\sqrt{N-1}}\,\sum_{k \neq j} |k\rangle\,.
\end{equation}
The initial state is characterized by $\vartheta = \vartheta_0$
with $\sin \vartheta_0 = 1/\sqrt{N}$. Each iteration transforms
$|\Psi(\vartheta)\rangle$ into $|\Psi(\vartheta+\omega)\rangle$,
where\, $\sin\omega = 2 \sqrt{N-1}/N$. Then, after $m \approx
(\pi/4)\sqrt{N}$ iterations, $\vartheta$ becomes very close to
$\pi/2$, and a measurement of the state yields $|j\rangle$ with an
error $O(1/N)$. We note that the evolution of
$|\Psi(\vartheta)\rangle$ according to the GA is restricted to a
2-dimensional H{\sc{ILBERT}} space which is spanned by $|x\rangle
= (1/\sqrt{N-1}) \sum_{k \neq j} |k\rangle$ and $|y\rangle =
|j\rangle$. Each iteration represents a rotation of the quantum
state by the angle $\omega$ in the $x$-$y$ plane and the
G{\sc{ROVER}}'s operator for a single iteration can be written in
a familiar form
\begin{equation}
\hat{R}(\omega)\; =\; \left(
\begin{array}{lcr}
\cos \omega && -\sin \omega\\
\sin \omega && \cos \omega
\end{array}
\right)
\end{equation}
on the basis $\left\{|x\rangle, |y\rangle\right\}$.

Imperfections are introduced in the GA as follows: the ideal
H{\sc{ADAMARD}} operator in the step~(ii) is given by $\vec{n}
\cdot \vec{\hat{\sigma}}$, where $\vec{n} = (1/\sqrt{2}, 0,
1/\sqrt{2})$, and $\hat{\sigma}_{x(y,z)}$ denotes the P{\sc{AULI}}
spin matrix. We now replace $\vec{n}$\, by
\begin{equation}
\vec{m}_q\; =\; \frac{1}{\sqrt{2}}\, (\cos \varphi_q \cdot \sin
\delta_q + \cos \delta_q ,\, \sqrt{2}\,\sin \varphi_q \cdot \sin
\delta_q ,\, -\cos \varphi_q \cdot \sin \delta_q + \cos
\delta_q)\,,
\end{equation}
where $q = 1,2, \cdots, n_q$ represent each single qubit. Here,
$\delta_q$ and $\varphi_q$ with $|\delta_q| < \epsilon/2$ and $0
\leq \varphi_q < 2\pi$ are randomly chosen in an iteration of the
GA and also vary randomly from iteration to iteration. Then, it
turns out that $\vec{m}_q$ is a unit vector tilted from $\vec{n}$
by $\sim \epsilon$. It should be noted that, in spite of the
imperfections, since the quantum state evolves without coupling to
the additional environment, the qubit rotations remain unitary,
keeping the normalization condition $\langle \Psi |\Psi \rangle =
1$ for any iteration number $t$. The presence of the imperfections
will provide an additional coupling between the 2-dimensional
H{\sc{ILBERT}} space spanned by $\left\{|x\rangle,
|y\rangle\right\}$ (``computational space'') and the rest part of
the total H{\sc{ILBERT}} space with $2^{n_q}$ dimensions, leading
to the {\em quantum leakage}\cite{FAZ99} from the computational
space as an intrinsic source of error in ideal gate operations.

Typical results of the GA with imperfections are shown in Fig.~1:
$\langle p_j\rangle$ and $F$ denote an ensemble-averaged
probability of the target state $|j\rangle$ and an
ensemble-averaged fidelity over 100 random runs, respectively,
each of which is here given for $n_q =13$ and for imperfection
strengths $\epsilon = 0$, 0.005, 0.01 and 0.02, respectively, as a
function of iteration number $t$. Clearly, they are given by
\begin{equation}
\langle p_j\rangle_{\epsilon}(t)\; =\; \left\langle |\langle
j|\Psi(\epsilon,t)\rangle |^2 \right\rangle\;,\;\;
F_{\epsilon}(t)\; =\; \left\langle |\langle
\Psi(\epsilon=0,t)|\Psi(\epsilon,t)\rangle |^2 \right\rangle\,,
\end{equation}
respectively, where the outer bracket represents the ensemble
average. In the case of $\epsilon=0$, $\langle p_j\rangle_{0}(t)$
oscillates between 0 and 1 and reaches 1 at $t \approx (m+1/2)
(\pi/2)\sqrt{N} \approx 71,\,213,\,355, \cdots$ with $m=0,1,2,
\cdots$. When $\epsilon$ is non-zero, one still finds oscillating
features with the same period as in the ideal case, however, with
an envelope decaying nearly exponentially with time $t$. As $t$
increases, the system approaches a saturated regime, where the
noise completely dominates the ideal system dynamics, and
accordingly $\langle p_j\rangle_{\epsilon}(t)$ fluctuates around
$1/N$. A novel feature is that the decay affects not only the
shape of the upper envelope but also that of the lower envelope
such that the lower envelope is not simply given by $\langle
p_j\rangle_{\epsilon}(t) = 0$. This means that the probability for
the system to remain at the target state is still available even
at the time it originally vanishes in the ideal unperturbed
system. Furthermore, $F_{\epsilon}(t)$ is found to approximately
equal the upper envelope of $\langle p_j\rangle_{\epsilon}(t)$.

As noted earlier, in the absence of imperfections, the
wave-function of quantum register evolves within a very small part
(of dimension 2) of the total H{\sc{ILBERT}} space (of dimension
$2^{n_q}$). Furthermore, since the amplitudes of $|x\rangle$ and
$|y\rangle$ remain real or at least keep the same phase over the
time evolution, the actual relevant space is even smaller than the
2-dimensional entire H{\sc{ILBERT}} space. Let us denote the
2-dimensional H{\sc{ILBERT}} space spanned by $|x\rangle$ and
$|y\rangle$ and the total H{\sc{ILBERT}} space by ${\cal H}_2$ and
${\cal H}_t$, respectively. The above results suggest that in
general, the disordered GA operator yields states which are not
restricted in ${\cal H}_2$ but spread over a larger space ${\cal
H}_t$ (``computational leakage''). In other words, the presence of
imperfections induces a probability density flow from ${\cal H}_2$
to ${\cal H}_t$ with diffusion-like nature. Then, let us define
$|w_2(t)|^2$ as the probability that the state remains in ${\cal
H}_2$ at time $t$ with an exponentially decaying function of $t$,
\begin{equation}
w_2(t) = e^{-\gamma\,t}\,,
\end{equation}
where $\gamma$ represents the strength of the diffusion which
depends on system parameters such as the strength of imperfections
and the qubit numbers. Also, the imperfections affect the dynamics
of the state within ${\cal H}_2$: in general, the phases of the
two amplitudes of $|x\rangle$ and $|y\rangle$ are not equal to
each other, and it is reasonable to assume that random phases are
introduced during each iteration. Therefore, we would now like to
adopt an effective two-level model which can encapsulate the
effects of imperfections in the GA living in ${\cal H}_t$. Here,
the time evolution of a quantum state $|\psi(t)\rangle =
c_m(t)\,|m\rangle + c_n(t)\,|n\rangle$ on the bais $\{|m\rangle,
|n\rangle\}$ is described by
\begin{equation}\label{eq:c_iteration}
\left(
\begin{array}{c}
c_m(t+1)\\
c_n(t+1)
\end{array}
\right)\; =\; e^{-\gamma}\; \hat{R}(\omega)\;
\hat{U}(\phi_m,\phi_n)\; \left(
\begin{array}{c}
c_m(t)\\
c_n(t)
\end{array}
\right)\,,
\end{equation}
where $\hat{U}(\phi_m,\phi_n)$ is a diagonal matrix with $U_{mm} =
e^{i\phi_m}$ and $U_{nn} = e^{i\phi_n}$, and $\phi_{m}(t)$ and
$\phi_{n}(t)$ are assumed to be two independent random variables
without any time correlation. Let each of these phase variables be
chosen from a box distribution $[-W_{\phi}/2, W_{\phi}/2]$ for a
given $W_{\phi}$. The frequency $\omega =
\sin^{-1}(2\sqrt{N-1}/N)$ is the same as in the GA, and the
initial conditions are given by $c_m(0) = \cos \vartheta_0$ and
$c_n(0) = \sin \vartheta_0$ with $\vartheta_0 =
\sin^{-1}(1/\sqrt{N})$. This is a stochastic two-level model with
dissipation and we refer to it as STLM hereafter. Here, we obtain,
after a minor calculation, an ensemble-averaged probability of the
target state $|j\rangle$ and an ensemble-averaged fidelity,
respectively:
\begin{equation}\label{eq:stlm}
\langle p_j\rangle_{W_{\phi}}^{(\gamma)}(t)\; =\; \left\langle
|c_n(t)|^2 \right\rangle\,,\; F_{W_{\phi}}^{(\gamma)}(t)\; =\;
\left\langle |c_m(t)\, \cos(\omega t + \vartheta_0)\, +\, c_n(t)\,
\sin(\omega t + \vartheta_0)|^2 \right\rangle\,.
\end{equation}

It is noteworthy to consider the difference between the STLM and
the original GA with imperfections. First, the finite fraction
$2^{-n_q+1}$ occupied by ${\cal H}_2$ in ${\cal H}_t$ is neglected
in the STLM so that $w_2(t)$ decays to zero instead of $\sim
\sqrt{2^{-n_q+1}}$. Since we are interested mainly in the regime
before saturation, this is clearly not a significant difference.
Secondly, the stochastic features of $\gamma$ are not considered.
But, this is not critical, either, since those features will
contribute a negligible correction to $\gamma$ after an
ensemble-average in eq.~(\ref{eq:stlm}).

Now, we perform a numerical simulation to obtain $\langle
p_j\rangle_{W_{\phi}}^{(\gamma)}(t)$ and
$F_{W_{\phi}}^{(\gamma)}(t)$, which will be compared with $\langle
p_j\rangle_{\epsilon}(t)$ and $F_{\epsilon}(t)$ of the GA with
imperfections, respectively. In Fig.~1, the results from the STLM
are shown as solid lines: they are given by ensemble-averages over
$1000$ realizations, respectively. We find that these results from
the STLM provide an impressive agreement with the results of the
GA after a proper adjustment of $\gamma$ and $W_{\phi}$. This
suggests that the main physical ingredients of the disordered GA
are correctly incorporated in the STLM. Nevertheless, the origin
of the novel feature in the lower envelopes is still unclear.

Without loss of generality, $(c_m(t),c_n(t))$ in the STLM during
the time evolution can be written by $(e^{-\gamma\,t} \cos
\vartheta(t), e^{-\gamma\,t+i\,\phi(t)} \sin \vartheta(t))$ with
$\phi(t) := \phi_n - \phi_m$ up to an overall phase. In case of
$\phi(t) \equiv 0$ for arbitrary $t$, the angle $\vartheta(t)$
increases by $\omega$ after each iteration and is then given just
by $\omega\,t + \vartheta_0$. However, if $\phi(t)$ does not
vanish, then from its stochastic nature, it follows that
$\vartheta(t)- \vartheta(t-1)$ is not constant but would fluctuate
around $\omega$.\cite{com} Now, under the assumption that
$\vartheta(t)$ and $\phi(t)$ are not correlated with each other,
but simply two random variables, we can find analytic expressions
of $\langle p_j\rangle_{W_{\phi}}^{(\gamma)}(t)$ and
$F_{W_{\phi}}^{(\gamma)}(t)$, respectively; let $\vartheta(t)$
increase by $\omega + \eta_{t-1}$ between $t-1$ and $t$ such that
\begin{equation}
\vartheta(t)\; =\; \vartheta_0\, +\, \omega\,t\, +\,
\sum_{k=0}^{t-1}\,\eta_k\,,
\end{equation}
where $\eta_k$ forms a G{\sc{AUSS}}ian distribution with mean 0
and width $\Delta_{\vartheta}$, and then $\sum_{k=0}^{t-1} \eta_k$
also satisfies a G{\sc{AUSS}}ian distribution with mean $0$ and
width $\Delta_{\vartheta} \sqrt{t}$. From this and
eq.~(\ref{eq:stlm}), we get:
\begin{eqnarray}\label{eq:p_t}
\langle p_j\rangle^{(\gamma)}(t) &=& |w_2(t)|^2\, \left\langle
\sin^2 \vartheta(t) \right\rangle\; =\;
\frac{|w_2(t)|^2}{\Delta_{\vartheta} \sqrt{\pi t}}\,
\int^{\infty}_{-\infty} \sin^2(\omega t + \vartheta_0 + x)\;
e^{-x^2/(\Delta^2_{\vartheta}\,t)}\; dx\nonumber\\
&=& \frac{e^{-2\,\gamma\,t}}{2}\, \left[\,1 - \cos(2\,\omega\,t +
2\,\vartheta_0) \cdot e^{-\Delta_{\vartheta}^2\,t}\,\right]
\end{eqnarray}
(note that no subindex $W_{\phi}$ appears in $\langle
p_j\rangle^{(\gamma)}(t)$\,). If we further assume that $\phi(t)$
also is of a G{\sc{AUSS}}ian distribution with mean 0 and width
$\Delta_{\phi}$, we then arrive at
\begin{equation}\label{eq:F}
F^{(\gamma)}(t)\; =\; \frac{e^{-2\,\gamma\,t}}{2}\, \left[\,1 +
e^{-\Delta_{\vartheta}^2\,t}\,\left\{1 - \sin^2(2\,\omega\,t +
2\,\vartheta_0) \cdot \left(1 -
e^{-\Delta_{\phi}^2/4}\right)\,\right\}\,\right]\,.
\end{equation}
Fig.~2 shows a comparison between the results of the GA with
imperfections and those of eqs.~(\ref{eq:p_t}) and (\ref{eq:F}).
The good agreement in $\langle p_j \rangle(t)$ would provide an
explanation of why its lower envelope is not simply given by
$\langle p_j \rangle = 0$ in the GA with imperfections; the
uncertainty in the rotation angle during a single iteration
accumulates as the iteration proceeds. Then, $\vartheta(t)$ does
not represent a definite direction on a 2-dimensional plane but
spreads over an interval range $(-\Delta_{\vartheta} \sqrt{t},\,
\Delta_{\vartheta} \sqrt{t})$. This offers an additional decay
channel into the target state $|j \rangle$ after
ensemble-averaging (see the term $e^{-\Delta_{\vartheta}^2\,t}$ in
eq.~(\ref{eq:p_t})). Also, in eq.~(\ref{eq:F}) with $\Delta_{\phi}
= 0$ we have $F^{(\gamma)}(t) = (e^{-2\,\gamma\,t}/2)\left(1 +
e^{-\Delta_{\vartheta}^2\,t}\right)$, which is given by the solid
lines in Fig.~2 as the best fit of $F_{\epsilon}(t)$ of the GA
with imperfections. From eq.~(\ref{eq:F}), it immediately follows
that $F(\Delta_{\phi} \neq 0)$ is always less than
$F(\Delta_{\phi} = 0)$.

In summary, we have investigated imperfection effects on the time
evolution of the G{\sc{ROVER}}'s algorithm both numerically and
analytically. An effective two-level model with dissipation and
randomness has been introduced and the results show a good
agreement with the simulation results of the disordered
G{\sc{ROVER}}'s algorithm. It turns out that the main features in
the results of the disordered G{\sc{ROVER}}'s algorithm can be
understood through the diffusion-like behavior of quantum states
from the original partial H{\sc{ILBERT}} space into the full
H{\sc{ILBERT}} space. The two main decaying mechanisms found in
this work are its direct manifestations. Our finding will provide
a useful basis for study of more general imperfection effects in
quantum algorithms.

We would like to thank Prof. G. J. Iafrate (NC State Univ.) and
Prof. G. Mahler (Univ. of Stuttgart) for critical reading of the
manuscript.

\newpage
\begin{figure}
\epsfxsize=3.2in \epsfysize=5in \epsffile{fig1.epsi}
\caption{Behaviors of $\langle p_j\rangle\,(\bullet)$ and
$F\,(\circ)$ in the Grover's algorithm for the qubit number
$n_q=13$ with the imperfection strength (a) $\epsilon=0.005$, (b)
$\epsilon=0.01$ and $\epsilon=0.02$ (inset). Each data point is
given, for every 20 iterations, by an ensemble-average over 100
realizations. The dotted line of (a) represents $\langle
p_j\rangle$ for $n_q=13$ in the ideal case ($\epsilon = 0$). The
solid lines result from the stochastic two-level model described
in the text with parameters (a) $\gamma = 7.6 \times 10^{-4}$, (b)
$3.0 \times 10^{-3}$ and $1.3 \times 10^{-2}$ (inset), and (a)
$W_{\phi} = 0.089$, (b) 0.19 and 0.25 (inset), respectively.}
\label{fig1}
\end{figure}

\begin{figure}
\epsfxsize=3.2in \epsfysize=5in \epsffile{fig2.epsi}
\caption{Comparison between the results of the Grover's algorithm
with imperfections and the theoretical predictions given by
eqs.~(\ref{eq:p_t}) and (\ref{eq:F}). The symbols indicate the
same data as in Fig.~1. The parameters $\gamma$'s are the same as
in Fig.~1 with (a) $\Delta_{\vartheta} = 2.0 \times 10^{-2}$, (b)
$4.2 \times 10^{-2}$ and\, $3.5 \times 10^{-2}$ (inset),
respectively, and $\Delta_{\phi} = 0$ for all three.} \label{fig2}
\end{figure}


\begin{thebibliography}{1}
\bibitem{com}
After some calculations from eq.~(\ref{eq:c_iteration}), we
actually obtain $\vartheta(t+1) = \frac{1}{2}
\cos^{-1}[\cos(2\omega + 2\vartheta(t)) - \sin(2\omega) \cdot
\sin(2\vartheta(t)) \cdot (\cos\phi(t) - 1)]$, which is, for
general cases, clearly different from $\vartheta(t+1) =
\vartheta(t) + \omega$.
\end{thebibliography}
\end{document}